\documentclass[aps,pra,reprint,superscriptaddress,showkeys,amsmath,amssymb,longbibliography, nofootinbib]{revtex4-1}

\usepackage[english]{babel}
\usepackage{graphicx}
\usepackage{dcolumn}
\usepackage{bm}
\usepackage{cellspace}%
\usepackage{array}%
\usepackage{
	xcolor,
	xspace,
	siunitx,
	braket,
    ragged2e,
    boldline
}

\usepackage{hyperref}
\usepackage[xspace]{ellipsis}

\makeatletter    
\newcommand*{\balancecolsandclearpage}{%
  \close@column@grid
  \clearpage
  \twocolumngrid
}
\makeatother

\DeclareSIUnit\gauss{G}
\DeclareSIUnit\photons{photons}
\DeclareSIUnit\atoms{atoms}
\begin{document}

\title{
 Geometry of turbulent dissipation and the Navier-Stokes regularity problem
}

\author{Janet Rafner}
\affiliation{ScienceAtHome, Center for Hybrid Intelligence, Department of Physics and Astronomy, Aarhus University, Aarhus, Denmark}
\author{Zoran Gruji\'c}
\affiliation{Department of Mathematics, University of Virginia, Charlottesville, Virginia, United States of America}
\author{Christian Bach}
\affiliation{ScienceAtHome, Center for Hybrid Intelligence, Department of Physics and Astronomy, Aarhus University, Aarhus, Denmark}
\author{Jakob Andreas B\ae rentzen}
\affiliation{Department of Computer Graphics, Technical University of Denmark, Lyngby,  Denmark}
\author{Bo Gervang}
\affiliation{Department of Engineering, Aarhus University, Aarhus, Denmark}
\author{Ruo Jia}
\affiliation{Stanford University, Stanford, California, United States of America}
\author{Scott Leinweber}
\affiliation{Apple INC., San Francisco, California, United States of America}
\author{Marek Misztal} 
\affiliation{3Shape TRIOS A/S, Copenhagen, Denmark}
\author{Jacob Sherson}
\affiliation{ScienceAtHome, Center for Hybrid Intelligence, Department of Physics and Astronomy, Aarhus University, Aarhus, Denmark}

\date{\today}


\begin{abstract}
The question of whether a singularity can form in an initially regular
flow, described by the 3D incompressible Navier-Stokes (NS) equations, is a 
fundamental problem in mathematical physics. The NS regularity problem is
super-critical, i.e., there is a `scaling gap' between what can be established by
mathematical analysis and what is needed to rule out a singularity. A recently introduced
mathematical framework--based on a suitably defined `scale of sparseness' of the regions
of intense vorticity--brought the first scaling reduction of the NS super-criticality
since the 1960s. Here, we put this framework to the first numerical test using a 
spatially highly resolved computational simulation performed near a `burst' of the vorticity magnitude. 
The results confirm that the scale is well suited to detect the onset of dissipation and provide 
strong numerical evidence that ongoing mathematical efforts may succeed in closing the scaling gap, 
and arriving at criticality.
\end{abstract}

\maketitle

\section{Introduction}

Humans have been fascinated with the geometry of fluid flows for centuries. Well-known examples of artistic 
renditions of `coherent vortex structures' appearing in nature include the woodblock print \emph{Great Wave off Kanagawa} by Katsushika Hokusai, Vincent Van Gogh's \emph{Starry Night} and Leonardo Da Vinci's drawings of eddy motion. Da Vinci's studies were in fact among the first scientific studies of turbulent motion which then assumed a more rigorous form in the fundamental works of A.N. Kolmogorov \cite{Ko41-1, Ko41-2, Ko41-3}, L. Onsager \cite{On45, On49}, G.I. Taylor \cite{Ta35, Tay37} 
and others in the first half of the 20th century. One of the cornerstones of the turbulence phenomenology is the concept of the energy cascade: there is a nonlinear transfer of energy from larger to smaller scales; once a sufficiently small
scale is reached, the diffusion takes over and the energy starts to dissipate in the form of heat. 
Despite much progress over the decades, a complete, rigorous theory of turbulence remains 
elusive; in particular, the question as to the role that coherent vortex structures play in the theory of
turbulent dissipation.

\medskip

A flow of 3D incompressible, viscous, Newtonian fluid is described by the 3D Navier-Stokes equations (NSE), 
\[
 \frac{\partial \pmb{u}}{\partial t} + (\pmb{u}\cdot \nabla)\pmb{u}=-\nabla p + \nu \triangle \pmb{u} + \pmb{f},
\]
supplemented with the incompressibility condition $ \, \nabla \cdot \,
\pmb{u} = 0$, where $\pmb{u}$ is the velocity of the fluid, $p$ is the kinematic 
pressure, $\nu$ is the kinematic viscosity, and $\pmb{f}$ the external force.
\linebreak Taking the curl yields the vorticity formulation,
\[
  \frac{\partial \pmb{\omega}}{\partial t} + (\pmb{u} \cdot \nabla) \pmb{\omega} =  \nu \triangle \pmb{\omega} + 
  (\pmb{\omega} \cdot \nabla) \pmb{u} +  \, \nabla \times \pmb{f}
\]
where $\pmb{\omega}$ is the vorticity of the fluid, $\pmb{\omega} = \, \nabla \times \pmb{u} \, $.
The left-hand side represents the transport of the vorticity
by the velocity, the first term on the right-hand side generates the diffusion, and the second one is the vortex-stretching term;
these are the three principal physical mechanisms in the system. Despite its apparent simplicity, the mathematical theory of the 3D NSE 
is fundamentally incomplete. In particular, since the pioneering work of J. Leray in the 1930s \cite{Le34}, the question of whether a finite-time singularity can form in an initially regular (smooth) unforced flow is still open, and is one of the remaining
\emph{Millennium Prize Problems} put forth by the Clay Mathematics Institute (customarily referred to as `the
NS regularity problem'). Note that--in this context--a temporal point $T^*$ is a \emph{singularity} if the solution is
regular on some time-interval $(T^*-\epsilon, T^*)$ and the
limit of the maximum of the velocity (or--equivalently--the vorticity) magnitude is infinite as the time variable approaches $T^*$.
Such singularity could potentially form in either of the following two cases i) in Eulerian dynamics (zero viscosity), where a formation of ever smaller scales--paired with the formation of ever larger velocity gradients could continue indefinitely or ii) in the viscous case, if the formation of ever smaller scales stays above the threshold of the dissipation scale (shrinking to zero slower than the dissipation scale, as the flow approaches the singularity).

\medskip

The NS regularity problem is \emph{supercritical}; in other words there is a `scaling gap' between any
known regularity criterion and the corresponding \emph{a priori} bound. Here  a `regularity criterion' refers to
an analytic or geometric property of the solution sufficient to rule out a singularity, while
an `\emph{a priori} bound' refers to an analytic or geometric property of the solution that can be rigorously derived
from the NSE. Moreover, since the fundamental independent works of Ladyzhenskaya, Prodi and Serrin, as well as Kato and Fujita, in the 1960s \cite{La67, P59, S62, FK64}, no one has improved upon the regularity criteria, with respect to the intrinsic NS scaling transformation, and all the
\emph{a priori} bounds have been on the scaling-level of Leray's original energy bound. 
Recently, however, a new mathematical framework
\cite{BrFaGr18} enabled the first 
\emph{scaling reduction} of the NS super-criticality (more precisely, the \emph{a priori} bound derived 
represents a $40\%$ improvement over the energy level bounds) 
since the 1960s; the key concept in the theory is a
suitably defined `scale of sparseness' (a precise definition resides in the following section) of the super-level sets of the positive and the negative parts of the vorticity
components. At this point, a natural question arises of whether there might be an intrinsic obstruction to advancing
this method, or whether this new framework might have more to offer (a further reduction of the scaling gap).

In this work we present the first numerical analysis of this novel geometric scale applied to two conventional scenarios i) a Kida-vortex initialized simulation: a model flow for the study of possible singular events in the 3D incompressible fluid flows (away from the boundary), in particular because it exhibits a `burst' of vorticity maximum, resembling a finite singularity \cite{BP94} ii) a simulation in the realm of fully developed, homogeneous, isotropic turbulence  \footnote{One should note that homogenous and isotropic refers to the velocity description; the vorticity description is inherently inhomogeneous and anisotropic.}. 
The main result of this paper demonstrated that in the former case the scale of sparseness is not only actualized well beyond the guaranteed 
\emph{a priori} bound \cite{BrFaGr18}, but also just beyond the critical
bound \cite{BrFaGr18} sufficient for the diffusion to fully 
engage (overpowering the nonlinearity) and 
prevent the further growth of the vorticity magnitude. This result provides numerical confirmation that the aforementioned mathematical
framework is capable of accurately detecting the onset of turbulent dissipation, and furnishes a necessary validation of
the current
theoretical efforts in closing the scaling gap in the NS regularity problem within the framework. In addition, we provide arguments that shed new light on some of the classical work in the computational 
simulations of \emph{intermittent events} in 
turbulence.  More specifically, based on the theory presented in \cite{BrFaGr18} we provide--for the first time--a rigorous explanation for the eventual `slumps' in the `bursts' 
of the vorticity magnitude observed in computational modeling of the Kida vortex-initialized flows \cite{KM86, KM87, BP94}.

\medskip

In the second scenario, we analyze data sampled from the forced isotropic turbulence simulation publicly available from the Johns Hopkins Turbulence Data Base (JHTDB). This data set had previously only been analyzed for spectral scales, and this work presents the first analysis of the geometric properties of the simulated flow. Since the scale of sparseness is a \emph{small scale} (e.g., in the case of an ensemble of filamentary `regions of intense vorticity' (RIVs), it is comparable
to the transversal scale, i.e., to the diameter of an RIV, and not to the longitudinal scale, i.e., to
the length of an RIV), we expect it to exhibit a scaling trend consistent with being in the dissipation range. 
This was indeed confirmed by performing time series analysis 
on the data sampled from JHTDB. The results are presented in the Appendix.

\medskip

Finally, as we believe the general concept of the `scale of sparseness' could be potentially impactful in many other situations in science and engineering, it is meaningful to take a closer look at how to optimize the data analysis algorithms. In the absence of heuristics, all RIVs must be investigated and an exhaustive search for the maximum $r$ value conducted ($r$ denotes a scale comparable to the scale of sparseness). This brute force approach is computationally demanding. One field that seeks to systematize the search for heuristics is Computational Citizen science. \footnote{Citizen science based on a research problem with a complete mathematical description and a unique and easily calculable score/quality for each user solution (‘easy validation’), which enables a direct comparison between algorithmic performance and the human computation. This includes both highly complex (multi-dimensional) natural science research problem and artificially constructed social science research problems aimed at a detailed mathematical investigation of individual and collective problem solving. Human vs. algorithmic comparisons using black box optimization would also fall into the category of Computational Citizen science.}  Computational Citizen science  has proven a useful technique for optimization in other fields such as protein folding \cite{Cooper10}, RNA mapping \cite{Denny18} and quantum physics \cite{Heck18} \cite{Sorensen16}. Thus, we explored the use of a digital, gamified interface (seen in Figure 2) which allowed 700 citizen-scientists from around the world to participate in the study of the scale of sparseness. In the context of the Kida vortex-initialized simulation, quantitative analysis of players' searches shows that they were able to achieve a comparable level of accuracy in identifying the $r_{max}$ (which was visualized as the radius of the largest sphere that can fit inside each RIV; a scale comparable to the scale of sparseness). The results can be seen in Figure 4).

\medskip

The paper is organized as follows. Section 2 provides a mathematical background necessary to present our results
(in particular, the mathematical framework suitable for quantifying the phenomenon of 
`spatial intermittency' introduced in \cite{BrFaGr18}), Section 3 exposes the main results,
Section 4 contains the discussion and the outlook, and Section 5 delineates the methods utilized.

\section{Mathematical background}

A key role played by geometry of the flow had been announced at least as far back as G. I. Taylor's 
fundamental paper \emph{Production and dissipation 
of vorticity in a turbulent fluid} in 1930s. Taylor inferred his thoughts on turbulent dissipation
partly from the wind tunnel 
measurements of a turbulent flow past a uniform grid, concluding the paper with the following statement:
``It seems that the stretching of vortex filaments must be regarded as the principal mechanical cause of 
the high rate of dissipation which is associated with turbulent motion'' \cite{Tay37}.

\medskip

A pioneering event in incorporating geometry of the flow in the mathematical study of possible 
singularity formation in the 3D NSE took place in P. Constantin's paper 
\emph{Geometric statistics in turbulence} \cite{Co94}. This approach was based on a singular integral
representation for the stretching factor in the evolution of the vorticity magnitude, and the key 
observation was that the kernel had a geometric flavor; more precisely, it was \emph{depleted} by
the \emph{local coherence} of the \emph{vorticity direction} (this has been referred to as `geometric depletion
of the nonlinearity').
The first quantification of this phenomenon appeared in \cite{CoFe93}, revealing that postulating 
the Lipschitz-continuity
\footnote{Let $\Omega$ be a subset of $\mathbb{R}^3$. Recall that a function $\pmb{f} : \Omega \to \mathbb{R}^3$ is Lipschitz-continuous if there exists a constant $L$ such that 
$\|\pmb{f}(x)-\pmb{f}(y)\| \le L \|\pmb{x}-\pmb{y}\|$ for all $\pmb{x}$ and $\pmb{y}$ in $\Omega$. Complementary, $\pmb{f}$ is
H\"older-continuous with the exponent $\alpha$, $0 < \alpha < 1$, if there exists a constant $A$
such that $\|\pmb{f}(x)-\pmb{f}(y)\| \le A \|\pmb{x}-\pmb{y}\|^\alpha$ for all $\pmb{x}$ and $\pmb{y}$ in $\Omega$.}
of the vorticity
direction in the regions of intense vorticity suffices to guarantee that the flow initiated at a regular initial
configuration does not develop a singularity. 

Subsequent results include \cite{BdVBe02} where it was shown 
that the assumption on the Lipschitz-continuity
can be scaled down to $\frac{1}{2}$-H\"older-continuity, \cite{Gr09} in which
a complete \emph{spatiotemporal localization}
of the $\frac{1}{2}$-H\"older-continuity regularity criterion was performed, and \cite{GrGu10} in which the authors presented
a two-parameter family of local, hybrid
geometric-analytic, \emph{scaling-invariant} regularity criteria, based on a \emph{dynamic balance} between the coherence
of the vorticity direction and the vorticity magnitude. 

\medskip

Another (in addition to the local coherence of the vorticity direction) morphological signature of the regions
of intense vorticity is \emph{spatial intermittency}--this has been well-documented in computational 
simulations as well as experiments. Classical references on the morphology of turbulent flows include
\cite{AKKG87, JWSR93, S81, SJO91, VM94}.  One should also note that--in contrast to the more traditional approach
of chasing the super-high Reynolds number regimes--a relatively recent work
 \cite{SchSree} revealed that, in sufficiently resolved flows, the universality in the realm of the small-scale structures already 
transpires
at low to moderate Reynolds numbers.

\medskip

The concepts of local 1D and 3D `sparseness' 
introduced in \cite{Gr13}
were designed to model the spatial intermittency in a way amenable to rigorous mathematical analysis
based on the 3D NSE. Henceforth, sparseness will refer to 3D sparseness defined
below ($m^3$ denotes the 3-dimensional
Lebesgue measure).

\medskip

Let $S$ be an open subset of $\mathbb{R}^3$, $\delta \in (0,1)$, $\pmb{x}_0 \in \mathbb{R}^3$,
and $r \in (0, \infty)$. $S$ is {\it 3D $\delta$-sparse around $\pmb{x}_0$ at scale $r$} if 
\[
\frac{m^3\bigl(S\cap B(\pmb{x}_0,r)\bigr)}{m^3\bigl(B(\pmb{x}_0,r)\bigr)} \leq \delta.
\]

\medskip

The main idea presented in \cite{Gr13} was to utilize sparseness of the vorticity super-level sets via the
\emph{harmonic measure maximum principle} (for the essentials on the harmonic measure in the 
complex plane see, e.g., \cite{Ahl}); in short, the sparseness translates 
into smallness of the
harmonic measure associated with the regions of the intense vorticity which--in turn--provides a bound on the 
$L^\infty$-norm \footnote{ Let $f$ be a Lebesgue measurable, scalar or vector 
valued function on an open set $\Omega$ in the Euclidean space.
Recall that for an exponent $p$, $1 \le p < \infty$, 
the Lebesgue space of $p$-integrable functions on $\Omega$, denoted by $L^p$, is determined
by finiteness of the $L^p$-norm,
\[
 \|f\|_p = \biggl( \int_\Omega |f(x)|^p \, dx \biggr)^\frac{1}{p}.
\]
In the limit case $p=\infty$, the corresponding space of essentially bounded functions, $L^\infty$, is determined
by finiteness of the $L^\infty$-norm,
\[
 \|f\|_\infty = \sup_{x \in \Omega} |f(x)|
\]
(the supremum is taken almost everywhere-$x$ with respect to the Lebesgue measure).
Closely related are the `weak Lebesgue spaces' $L^p_w$ determined by the rate of decay of the
volume of the super-level sets of a function. Let $\lambda > 0$ and consider the super-level set of the function 
$f$ cut at the level $\lambda$,
\[
 S_\lambda = \{ x \in \Omega : \, |f(x)| > \lambda\}.
\]
The function $f$ belongs to $L^p_w$ if there exists a positive constant $c$ such that
\[
 m^3 \biggl(S_\lambda\biggr)  \le \frac{c^p}{\lambda^p}
\]
for all $\lambda > 0$. An $L^p$ function is automatically in $L^p_w$; the converse is false.}, 
controlling a possible finite time blow-up. A key PDE technique utilized here was deriving
the lower bounds
on the radius of spatial analyticity of the solution in the $L^\infty$-type spaces. The principal result obtained
in \cite{Gr13} states that as long as the suitably defined vorticity super-level sets are sparse at the scale comparable
to the radius of spatial analyticity measured in $L^\infty$, no finite time blow-up can occur. Since the
local-in-time spatially analytic smoothing is simply a (strong) manifestation of the diffusion, this result is
consistent with the physics of turbulent dissipation.

As noted in the Introduction, the Navier-Stokes regularity problem is \emph{supercritical}, i.e., there is a `scaling gap between any known regularity criterion and
the corresponding \emph{a priori} bound. If we focus on $L^\infty((0,T), X)$-type spaces ($L^\infty$ in time, spatially
in $X$),
an illustrative example is in the $L^p$-scale where the regularity class is
$L^\infty((0,T), L^3)$ \cite{ESS03}, and the
corresponding \emph{a priori} bound the Leray bound $L^\infty((0,T), L^2)$. Moreover, since the 
fundamental (independent) works of Ladyzhenskaya, Prodi and Serrin, as well as Kato and Fujita, all in 1960s, 
the scaling
gap has been of the \emph{fixed size} in the sense that all the regularity classes are (at best) 
\emph{scaling-invariant} (with respect to the intrinsic NS scaling) while--in contrast--all the known \emph{a priori} bounds are
on the scaling level of the Leray \emph{energy bound}.

It turned out that redesigning the theory introduced in \cite{Gr13} around the super-level sets of the \emph{vorticity components} instead of the vectorial super-level sets (see Figure 1) produced the first `modern era'
\emph{algebraic reduction} of the scaling gap; since 1960s, all the improvements were logarithmic in
nature, regardless of the mathematical framework utilized. 

\begin{figure}[t]
  \begin{tabular}{cc}

    \includegraphics[width=40mm]{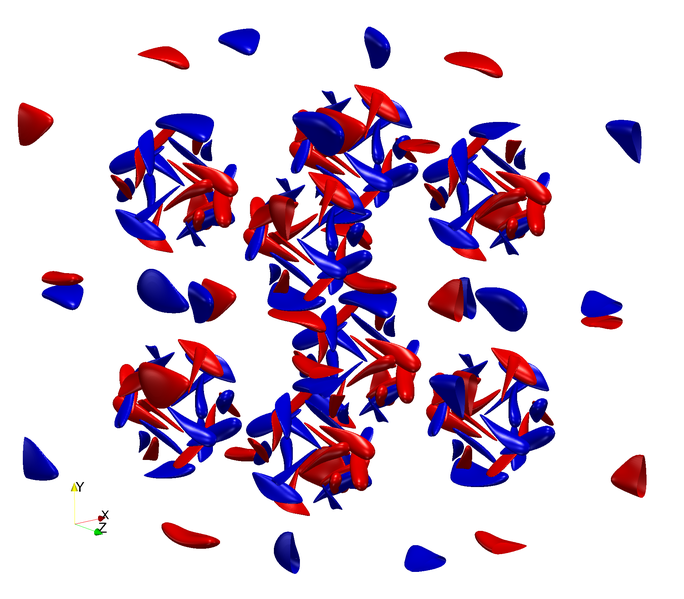}&

    \includegraphics[width=40mm]{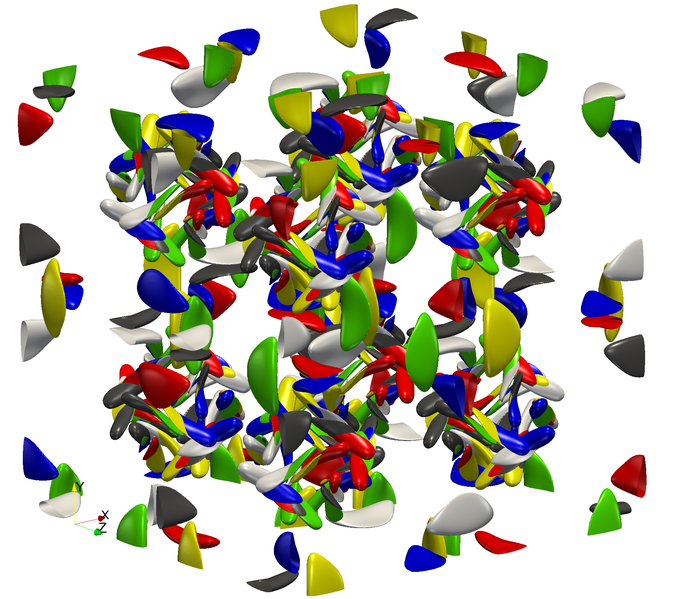}\\
  \end{tabular}
    \caption{An illustration of the difference between sparseness of the super-level sets of 
  the individual components and
  the full vectorial super-level set. The figure on the left depicts the super-level sets of the positive 
  and the negative 
  parts of the first component of the vorticity field, while the figure on the right--the union of the six individual
  super-level sets--corresponds to the super-level set of the vorticity magnitude. Obtained from the Kida simulation.}
    \label{fig:overview_figure}
\end{figure}

\begin{figure*}[t]

		\includegraphics[width=180mm]{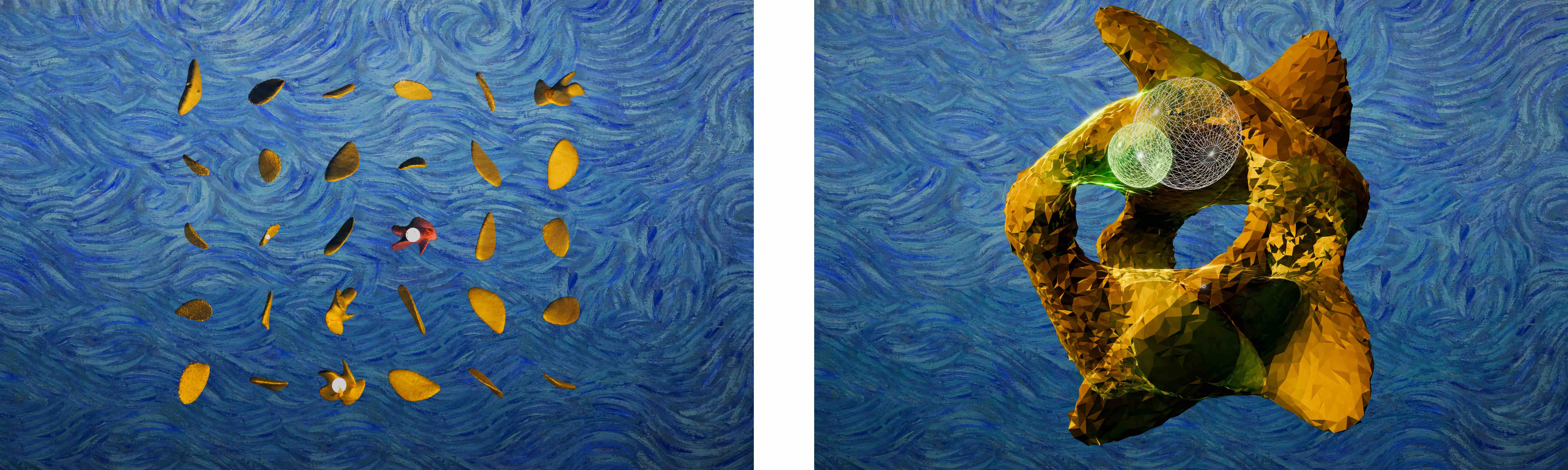}
		\caption{Screen captures from the Turbulence computational citizen science game showing first the selection of the RIVs and second, moving a sphere inside an individual RIV. Recall, the radius of the 
		largest sphere that can fit inside each individual RIV represents the scale of sparseness.}

\end{figure*}
\medskip

Details can be found in \cite{BrFaGr18}; here 
we present the classes $Z_\alpha$ as well as the synopsis of the two main results encoded in the 
$Z_\alpha$-formalism. Henceforth, for a vector field $\pmb{f}$, the positive and the opposite of the negative parts of
the components $f_i$ will be denoted by
$f_i^\pm$, $i=1, 2, 3$.

\medskip

For a positive exponent $\alpha$, and a selection of parameters $\lambda \in (0,1)$, $\delta \in (0,1)$
and $c_0>1$, the class of functions
$Z_\alpha(\lambda, \delta; c_0)$ consists of bounded, continuous functions
$\pmb{f} : \mathbb{R}^3 \to \mathbb{R}^3$ subjected to the following uniformly local condition. 
For $\pmb{x}_0 \in \mathbb{R}^3$,
select the/a component $f_i^\pm$ such that $f_i^\pm(\pmb{x}_0) = \|\pmb{f}(\pmb{x}_0)\|$
(here, the norm of a vector $\pmb{v}=(a, b, c)$, $\|\pmb{v}\|$, will be computed as $\max \{|a|, |b|, |c|\}$),
and require that the super-level set
\[
 \biggl\{ \pmb{x} \in \mathbb{R}^3: \, f_i^\pm(x) > \lambda \|\pmb{f}\|_\infty\biggr\}
\]
be $\delta$-sparse around $\pmb{x}_0$ at scale $\frac{1}{c} \frac{1}{\|\pmb{f}\|_\infty^\alpha}$,
for some $c, \frac{1}{c_0} \le c \le c_0$. Enforce this for all $\pmb{x}_0 \in \mathbb{R}^3$.
(In short, we require sparseness of the/a \emph{locally maximal} component only.)

\medskip

In this setting, the regularity class is $Z_\frac{1}{2}$ (pointwise-in-time, in the context of a blow-up
argument) \cite{BrFaGr18}; this is on the scaling level of
all the classical regularity criteria. The \emph{a priori} bound obtained lives in $Z_\frac{2}{5}$
(pointwise-in-time, in the context of a blow-up argument) \cite{BrFaGr18}; since the energy level class is
$Z_\frac{1}{3}$, this represents a $40\%$ reduction of the scaling gap in the $Z_\alpha$ framework!

\medskip

It is instructive
to make a quick comparison between the $Z_\alpha$-classes and the weak Lebesgue classes 
$L^p_w$ determined solely by the rate of decay of the volume of the super-level sets, 
encoding no geometric information. On one hand, it is straightforward that
$\displaystyle{
 f \in L^p_w \ \mbox{implies} \, f \in Z_\alpha \  \mbox{for} \  \alpha=\frac{p}{3}}$
(for a given selection of $\lambda$ and $\delta$, the tolerance parameter $c_0$ will depend
on the $L^p_w$-semi-norm of $f$);
on the other hand, in the geometrically worst case scenario--the whole super-level set being clumped into a 
single ball--being in $Z_\alpha$
is consistent with being in $L^{3 \alpha}_w$ (however, one should note that--in general--membership in 
$Z_\alpha$ does not impose any decay on the volume of the super-level sets since it provides information
on the suitably defined scale of the `largest piece' but no information on the number of `pieces'). 
This observation reveals that 
a `geometrically blind' 
scaling counterpart of the  \emph{a priori} bound in $Z_\frac{2}{5}$ would be the 
bound $\omega \in L^\infty((0,T), L^\frac{6}{5}_w)$ which is well beyond 
state-of-the-art ($\omega \in L^\infty((0,T), L^1)$ \cite{Co90}), demonstrating a clear advantage of working
within the $Z_\alpha$ realm compared to the functional classes traditionally utilized in the study of the
3D NS regularity problem.

\section{Results}

\subsection{Kida-vortex initialized flow: towards closing of the scaling gap}

Recall that in the $Z_\alpha$ framework, the rigorous regularity and \emph{a priori}-bound classes are
$Z_\frac{1}{2}$ and $Z_\frac{2}{5}$, respectively \cite{BrFaGr18}. In order to test whether there may be 
an obstruction to advancing the $Z_\alpha$ \emph{a priori} bounds even further--with the ultimate 
goal of \emph{closing} the
scaling gap--we considered a Kida-vortex initialized flow, a flow with the highly symmetric initial condition
exhibiting a sharp increase (a `burst') of the vorticity maximum `simulating' a finite-time singularity
\cite{KM86, KM87, BP94}.

\medskip

Attention was focused on a time interval leading to the peak of $\|\pmb{\omega}(t)\|_\infty$, and the aim was to
investigate a possibility of a power-law dependence between the actual geometric scale of sparseness $r(t)$
and the diffusion scale $d(t)=\frac{\nu^\frac{1}{2}}{\|\pmb{\omega}(t)\|_\infty^\frac{1}{2}}$ of the form $r \sim d^\alpha$
(recall that $d$ is a lower bound on the radius of spatial analyticity).
A presence of a power-law scaling would indicate that the scale of sparseness $r$ is a \emph{bona fide} 
small scale in the sense of turbulence phenomenology. In addition--since
both $d$ and $r$ are valued between 0 and 1--detecting a power law with an exponent 
$\alpha > 1$ would mean
that the solution in view is entering the diffusion regime (as depicted in the $Z_\alpha$ framework) and is, 
in fact, contained in the regularity 
class $Z_\frac{1}{2}$.

\medskip

Indeed, data analysis of the time-interval of interest-sourced from a spatially highly-resolved (${1024}^3$) simulation--revealed a very strong evidence of a power-law scaling; moreover,
the scaling exponent $\alpha$ crystalized at 1.098(9) (cf. Figure 3). This was very exciting to see, 
and the complementary rigorous analysis in the $Z_\alpha$ framework 
is currently underway (\cite{GrXu19}).

\subsection{Kida-vortex initialized flow: a rigorous explanation of the Boratav-Pelz computational results}

Boratav and Pelz considered the Kida vortex flows as--among other things--a laboratory for 
the computational study
of the possible singularity formation in solutions to the 3D NSE and Euler flows. In particular, they discovered a
time-interval of extreme intermittency (preceding the peak of the vorticity maximum) in which the local 
pointwise and integrated quantities increase sharply, noting that ``The increase is so sudden and sharp that
questions arise as to whether some of these quantities would diverge in finite time or not''  \cite{BP94}.

\medskip

They performed a singularity analysis revealing that the vorticity maximum--in a time-interval
leading to its peak--scales closely to the self-similar, critical scaling of $\frac{1}{t-T^*}$. This could, 
in principle, be 
an indication of the build-up of a scaling-invariant singularity. Nevertheless, the simulations consistently
showed an eventual disruption in the (approximately) self-similar, critical scaling, a formation of the peak, and a
subsequent dissipation of the flow, prompting them to conclude that ``However, the increase in peak vorticity
stops at a certain time, possibly due to viscous dissipation effects''  \cite{BP94}.

\medskip

As noted in the previous subsection, the analysis of the data sourced from an interval leading to the peak
of the vorticity maximum within the $Z_\alpha$ framework demonstrated that the flow entered a regime 
beyond the critical diffusion class $Z_\frac{1}{2}$. In other words, sparseness of the super-level
sets of the vorticity components was sufficient for the harmonic measure maximum principle to engage and
prevent a further growth of the vorticity maximum, providing a rigorous 
justification of the observed `slump' and dissipation. Nevertheless, it is worth noticing that the exponent observed \
in our analysis (1.098(9)) is only slightly larger than the critical exponent (1), confirming a close proximity to
the criticality.


\begin{figure*}[t]
	
	\centering
	
	\label{comp}\caption{The figure on the left is a Log-log plot of the scale of three dimensional sparseness (of the vorticity components) \emph{vs.} the diffusion scale. The summary of the regression is as follows: fit = $1.098(9) \cdot$ log (d) + 2.17(4). The inset compares the algorithmically calculated $r_{max}$ [the small dots] to the $r_{max}$ [the small circles] found by citizen scientists. The figure on the right is the time-evolution of the vorticity magnitude. The highlighted region indicates where the data was sourced from.}
	
	\begin{tabular}{cc}
		
		\includegraphics[width=85mm]{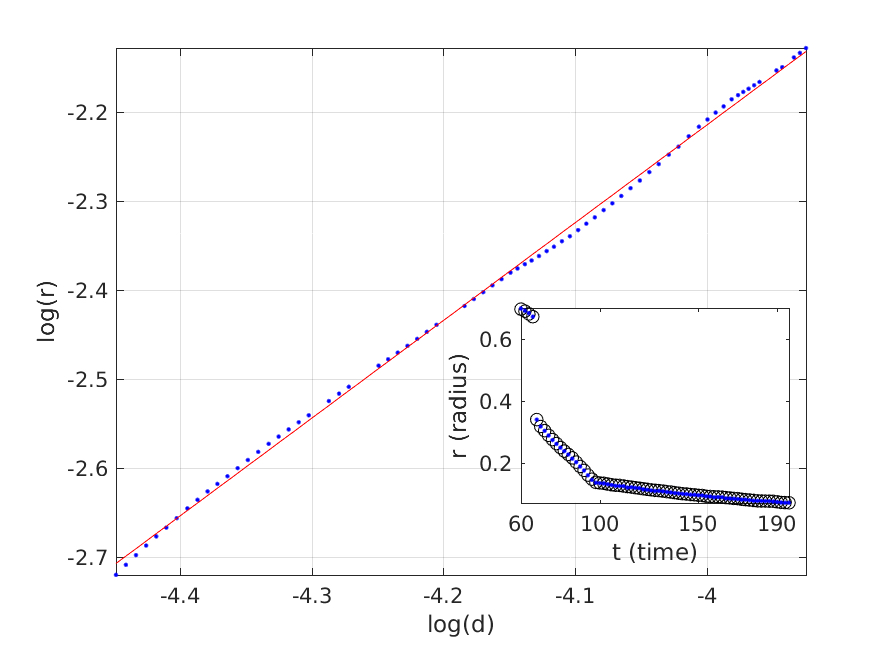}&
		
		\includegraphics[width=85mm]{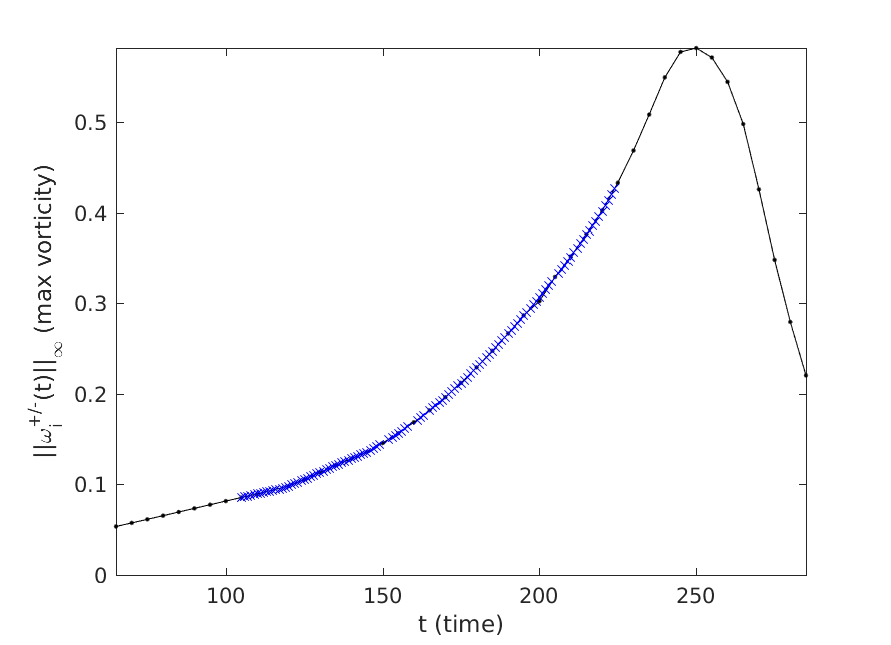}\\

	\end{tabular}
	
	 \label{fig3:figlabel}
	
\end{figure*}

\section{Discussion}

The main goal of this paper was to investigate whether there was an intrinsic obstruction to further
reduction of the scaling gap in the 3D Navier-Stokes regularity problem, within the mathematical
framework based on the suitably defined scale of sparseness of the regions of intense fluid activity. 
The idea was to perform a spatially highly resolved computational simulation of a Kida vortex-initialized
flow, and investigate the pertinent scaling properties as the flow
approached a peak of the vorticity magnitude. The data analysis exhibited that--in a time-interval leading
to the burst--the scaling signature of the scale of sparseness was 
slightly beyond the critical scaling (the scaling exponent of 1.098(9) \emph{vs.} 1), setting the 
stage--within the aforementioned framework--for 
the diffusion to fully engage and prevent further growth of the vorticity magnitude
(shortly, no roadblocks to criticality were detected).
To put this finding in perspective, on one hand, a scaling exponent less than 1 would indicate
that the new geometric 
framework does not offer a good gauge of whether the flow has entered a diffusion regime.
On the other hand, a scaling exponent significantly larger than one (i.e., significantly away from
the critical scaling) would not be consistent with the previously observed near-criticality features
of the Kida bursts (cf. \cite{BP94}). 

\medskip

The new geometric framework was designed to study the geometry of turbulent dissipation in the vicinity
of a hypothetical singularity (or a burst of the vorticity magnitude), regardless of the initial 
configuration. The Kida vortex-initialized flow was chosen as a notable example of a flow capable
of producing a sudden burst. It would be interesting to perform a similar analysis of other flows
with this capability (e.g., the flow studied in \cite{HK02}, featuring several peaks of the vorticity
magnitude, also initialized at an initial configuration concentrated on the first couple of Fourier
modes, but with no imposed symmetries).

\medskip 
The exploratory results from the computational citizen science game give first indications that it may be beneficial to study how an initial screening of human 'spot sorting' of these RIVs could potentially speed up our algorithmic processes.  This approach could be particularly useful when analyzing vast amounts of structures with comparable bounding box volumes, as this is the condition for reducing the number of RIVs analyzed with a distance field calculation. Apart from potential implications for analysis in computational fluid dynamics, this could offer significant insights within neurocomputation \cite{LB13} and computational geometry \cite{JBS6} while at the same time being an excellent outreach initiative to engage the general public in natural science research.

\medskip

The results exposed in this paper have provided a significant boost to the continuation of rigorous analysis, based on a countable hierarchy 
of function classes. Shortly, the original framework \cite{BrFaGr18} was based on sparseness 
of the super-level
sets of the vorticity, which could be viewed as the super-level sets of the first-order spatial fluctuation of the
velocity. It turns out that 
considering the super-level sets of the higher-order spatial fluctuations of the velocity field yields a further
reduction of the scaling gap. As a matter of fact, in a work just completed (\cite{GrXu19}), the authors succeeded in
establishing the \emph{asymptotic criticality} by showing the scaling gap shrinks to zero
as the order of the fluctuation goes to infinity. This, for the first time, revealed a critical nature of the
Navier-Stokes regularity problem.

\medskip

In conclusion, the numerical work presented here represents
the first-ever application of the abstract mathematical framework based on the concept of the `scale of sparseness' to concrete and realistic flows and our numerical success provides strong encouragement that this mathematical framework may provide novel insights in many realms of fluid dynamics in both science and engineering.

\section{Methods}

The Kida vortex initial configurations were introduced in \cite{K85}. The class features
a high number of symmetries, all preserved by the Navier-Stokes flows. The symmetries include
periodicity, bilateral symmetry, rotational symmetry and permutational symmetry of the velocity
components. More specifically, we considered the following one-parameter family of the initial
conditions,

 \begin{align*}
 u_{0,x} & = \sin x  \bigl( \cos (3y) \cos z - \cos y \cos (3z)\bigr) U_0\\
 u_{0,y} & = \sin y  \bigl( \cos (3z) \cos x - \cos z \cos (3x)\bigr) U_0\\
 u_{0,z} & = \sin z  \bigl( \cos (3x) \cos y - \cos x \cos (3y)\bigr) U_0
\end{align*}
(cf. \cite{BCK15}). The Reynolds number components were chosen as 
follows, $U_0 = 0.01,  L = 2 \pi$ ($2 \pi$-periodic box), $\nu = \frac{1}{3} 10^{-4}$,
resulting in the Reynolds number $Re = \frac{U_0 L}{\nu}$ of approximately $2000$.

\medskip

The data set was created utilizing a mixed spectral finite elements-Fourier method. The $XY$-plane was discretized with a Galerkin  pseudospectral discretization and extended in the z-direction with a Fourier discretization. This allows for a scalable solver and also exploits the speed of the Fast Fourier Transform similarly as in \cite{HKB7}. The Navier-Stokes equation was solved using an operator splitting method (Velocity Correction Scheme) leading to several systems of equations of reduced complexity to be solved instead of one large matrix system. That is while maintaining a splitting error on the order of the overall numerical discretization error. The temporal discretization was done with  Implicit-explicit (IMEX) schemes. Among other schemes, we specifically used an IMEX 3rd order scheme, which uses a third order Adams-Bashforth scheme for the convective term and a third order Adams-Moulton for the Diffusion term – The first order IMEX scheme analogously uses a combination of Euler schemes.

\medskip 

The data was analyzed mainly using a conventional algorithmic approach but also an exploratory citizen science approach described in the outlook. The algorithmic approach used a signed distance field calculation \cite{JBS6} where distances were computed from triangle mesh representations of the RIVs. A distance field for a surface is a mapping, $d: \mathbb{R}^3 \rightarrow \mathbb{R}$, from a point in space, $\mathbf p$, to the distance from $\mathbf p$ to its closest point on the surface. The sign of the distance field allows us to tell inside from outside, and the largest interior distance is also the radius of the largest sphere fully contained in the RIV. To compute the distance efficiently at a given point, $\mathbf p$, a bounding box hierarchy was used to find the triangles most likely to contain the point at a closest distance to $\mathbf p$. To compute the sign of the distance, we used ray casting to perform multiple tests for each point in order to determine if $\mathbf p$ were inside or outside the surface. This was necessitated by the fact that not all RIV meshes were orientable. On the other hand, significant optimizations were possible due to the fact that, for a given time slice, once an estimate of the greatest radius, $r$, had been found, we only needed to consider other RIVs if their bounding boxes were large enough to contain a sphere of radius $r$. Furthermore, we optimized the search for the largest radius within a given RIV by recursively computing distances at finer grid resolutions in the vicinity of the largest distance value found at the next coarser resolution.

\section{Appendix: Homogeneous, isotropic turbulence}

Here, we consider a forced isotropic
turbulence simulation made available through the JHTDB. The flow is kept in the statistical stationary 
state of the fully developed turbulence
by perpetually injecting the energy at large scales
(at low Fourier modes, effectively), keeping the kinetic energy
constant within the Fourier modes of the length less or equal to two. Any stationarity is exhibited
in the average only, and any pointwise-in-time-measured extremal quantities (e.g., the vorticity maximum,
$d$ and $r$) are expected to exhibit significant fluctuations. 

\medskip

As already noted in Introduction, the scale of sparseness is a small scale and--as such--is expected 
to trend as a dissipation range quantity.
In particular, in the framework of the $Z_\alpha$-classes, we expect it to stay within the confinements
of the critical diffusion/dissipation class $Z_\frac{1}{2}$. This was indeed confirmed via time series analysis of the data sampled from the JHTDB. More precisely, before filtering out the cyclic component, the scaling factor was 1.7(6), and after the filtering 1.5(2), with a tighter fit (cf. Figure 4). Recall that a scaling exponent greater than one corresponds to the dissipation range.

\begin{figure*}[t]
	
	\centering
	
	\label{comp}\caption{The figure on the left represents the linear regression after the cyclic component of the data was filtered out and the figure on the right the time-evolution of the vorticity magnitude. The summary of the regression is as follows fit = log(d) $\cdot$ 1.5(2) + 6.1(1.6).}
	
	\begin{tabular}{ccc}

		\includegraphics[width=85mm]{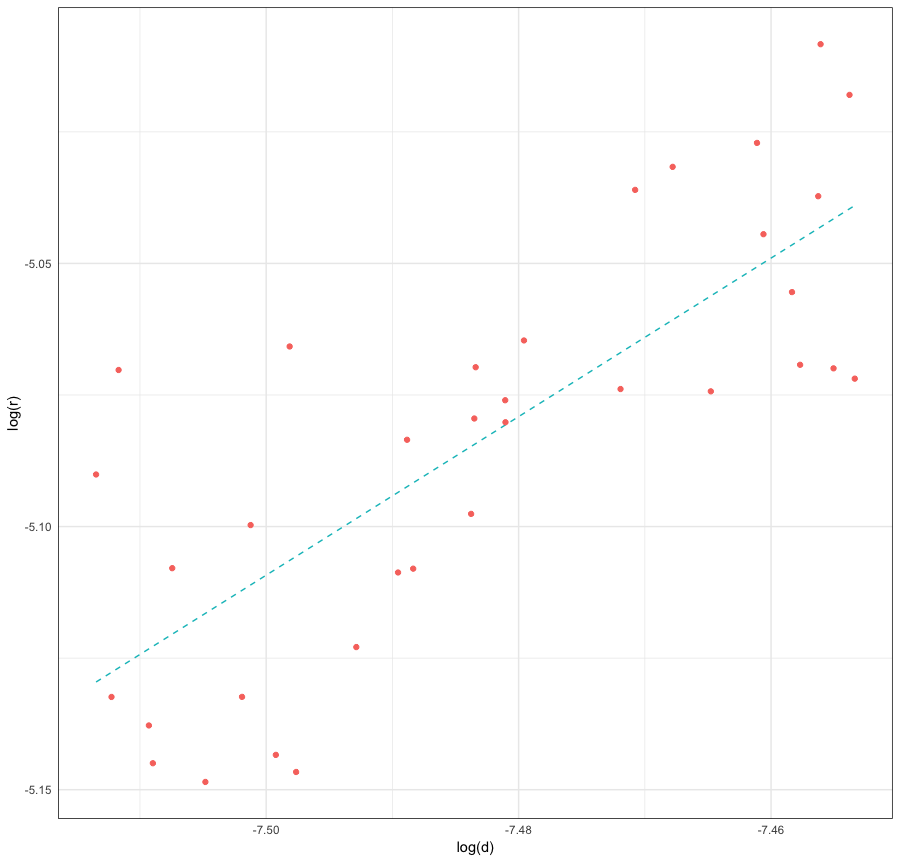}&
	    \includegraphics[width=85mm]{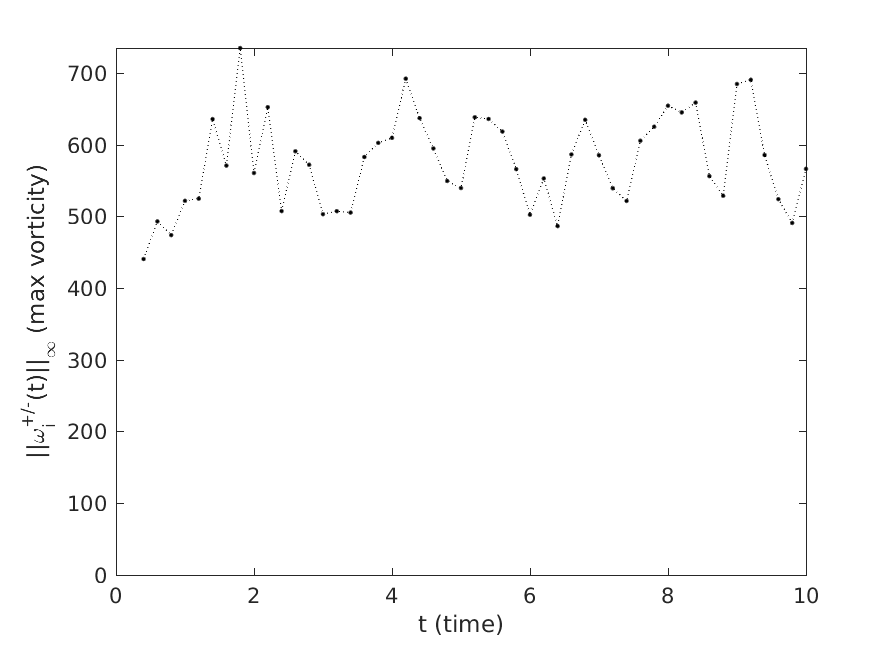}&
		
	\end{tabular}
	\label{fig4:figlabel}

\end{figure*}

\section{Acknowledgements}

We thank R. Feidenhans'l, S.\ Purup and J. Mathiesen for their input in the initial phases of this work. Zoran Gruji\'c gratefully acknowledges the financial support of the National Science Foundation (NSF). The Aarhus team thanks the Grundfos, John Templeton, Carlsberg, and Lundbeck Foundations for financial support. J.F.S. acknowledges funding from the ERC, H2020 grant 639560 (MECTRL). Finally, we thank all Turbulence Game players for participating in this study.

%

\end{document}